\documentclass{optica-article}

\journal{opticajournal} 

\articletype{Research Article}

\usepackage{lineno}

\begin{document}

\title{Coherent control of an optical tweezer phonon laser}

\author{K. Zhang,\authormark{1} K. Xiao,\authormark{2} D. Luntz-Martin,\authormark{3} P. Sun,\authormark{1} S. Sharma,\authormark{4} M. Bhattacharya,\authormark{2,5} and A. N. Vamivakas\authormark{1,3,5,6,*}}

\address{\authormark{1}The Institute of Optics, University of Rochester, Rochester, NY, USA\\
\authormark{2}School of Physics and Astronomy, Rochester Institute of Technology, Rochester, NY, USA\\
\authormark{3}Department of Physics and Astronomy, University of Rochester, Rochester, NY, USA\\
\authormark{4}Department of Physics, Korea Advanced Institute of Science and Technology, Daejeon 34141, South Korea\\
\authormark{5}Center for Coherence and Quantum Optics, University of Rochester, Rochester,NY, USA\\
\authormark{6}Materials Science Program, University of Rochester, Rochester, NY, USA}

\email{\authormark{*}nick.vamivakas@rochester.edu} 


\begin{abstract*} 
The creation and manipulation of coherence continues to capture the attention of scientists and engineers.  The optical laser is a canonical example of a system that, in principle, exhibits complete coherence.  Recent research has focused on the creation of coherent, laser-like states in other physical systems. The phonon laser is one example where it is possible to amplify self-sustained mechanical oscillations. A single mode phonon laser in a levitated optical tweezer has been demonstrated through appropriate balance of active feedback gain and damping.  In this work, coherent control of the dynamics of an optical tweezer phonon laser is used to share coherence between its different modes of oscillation, creating a multimode phonon laser.  The coupling of the modes is achieved by periodically rotating the asymmetric optical potential in the transverse focal plane of the trapping beam via trap laser polarization rotation. The presented theory and experiment demonstrate that coherence can be transferred across different modes of an optical tweezer phonon laser, and are a step toward using these systems for precision measurement and quantum information processing.

\end{abstract*}

\section{Introduction}
The generation, control, and sharing of coherence is ubiquitous across science and engineering \cite{mandel1995optical,glauber1963quantum}. The notion of a coherent state was formulated in the early days of quantum mechanics \cite{dirac1926theory,hall1994segal}. Shortly thereafter it was discovered that the self-sustained oscillation of first microwave \cite{pozar2011microwave,rohde2005design} and then optical radiation could occur when a gain medium is embedded in suitably designed cavity \cite{siegman1986lasers,saleh2019fundamentals}.  Both the maser and laser have had profound technological impact. Within the last twenty years, the same physics that gave rise to the maser and laser has been applied to mechanical systems. There have been extensive investigations into the creation of coherent mechanical motion; so called phonon lasers.  Demonstrations have occurred across diverse domains, ranging from atomic systems to microscale oscillators\cite{zhang2018phase,wei2022detection,wallentowitz1996vibrational,bargatin2003nanomechanical,vahala2009phonon,grudinin2010phonon,jing2014pt,wang2017demonstration,zhang2022dissipative,pettit2019optical,kuang2023nonlinear}. In this work, we focus on the implementation of a single mode phonon laser producing a coherent state using a levitated optomechanical system. \cite{pettit2019optical,kuang2023nonlinear}. 

As a mechanical oscillator, a levitated nanoparticle supports multiple degrees of freedom, including oscillation along three orthogonal directions, that exhibit thermal statistics.  Recent studies on multimode optomechanical  \cite{heinrich2011dynamics,mercade2021floquet,lee2015multimode,kohler2020simultaneous,wu2022generation} and electromechanical \cite{faust2013coherent,okamoto2013coherent,pernpeintner2018frequency} phonon lasers make it natural to realize a multimode optical tweezer phonon laser. One approach would necessitate the utilization of distinct feedback control techniques to induce coherent oscillation in each mode resulting in a complex system \cite{fang2016optical,pernpeintner2018frequency}.  Moreover, since the feedback signals are summed up to modulate the intensity of the trapping laser, extra noise may be injected resulting in uncontrolled ejection of the particle from the trap. 

An alternative approach to create multimode coherent states is to use dynamic coherent coupling.  In such an approach, with one mode prepared in a coherent state, the coherent coupling could share and/or transfer coherence between the coherent and thermal modes.  The net result of the coupling is to distribute coherence to the different phonon modes.  One route to realize such coupling is to periodically rotate the asymmetric trapping potential in the focal plane by modulating the trapping laser's polarization orientation \cite{frimmer2014classical}.  This coupling has been used to sympathetically cool the mean phonon number of a levitated nanoparticle \cite{frimmer2016cooling}. 

In this paper, we demonstrate a two mode phonon laser in levitated nanoparticle system, in which, the $x$ mode is prepared in a coherent state by simultaneously applying nonlinear feedback cooling and linear feedback heating. The $y$ mode, initially in a thermal state, is then driven into a coherent state via coherent coupling. We analyze the steady state dynamics to confirm the change from thermal state to coherent state as well as transient evolution of phonon number in two modes, which is well matched with our theoretical model.

\section{Method}

\begin{figure}
\centering
\includegraphics[width=0.7\textwidth]{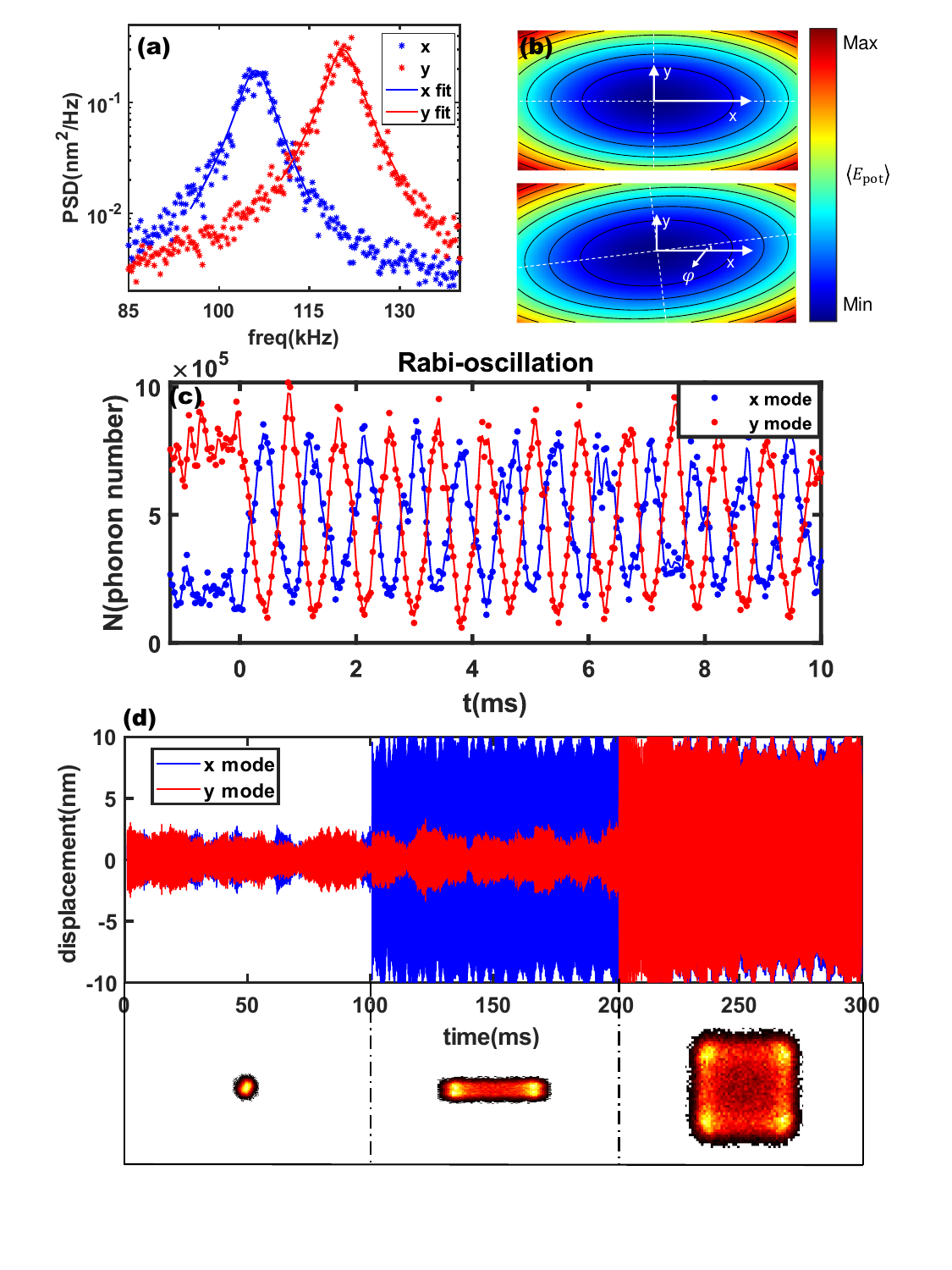}
\caption{\label{fig:one}Coupling between two oscillation modes.(a) Position spectral density (PSD) for the $x$ and $y$ oscillation modes at pressure of 6 mbar. (b)Top: Illustration of the optical trapping potential when the trapping beam is linearly polarized in the $x$ direction. Bottom: Asymmetric trapping potential rotated by an angle $\phi$ around the optical axis. (c) Rabi-oscillation between the two oscillation modes. Feedback cooling on the $x$ mode is on before t = 0 ms to prepare the $x$ mode in a lower energy state. Feedback cooling on $x$ is turned off and the coupling is switched on at t=0 ms.(d) Top: 300 ms time trace showing the center of mass (COM) displacement evolution. 0-100 ms: both the $x$ and $y$ modes are cooled. 100-200 ms: Amplification on the $x$ mode is above the threshold. 200-300 ms: coupling between the $x$ and $y$ mode is turned on. Below: Density distribution maps of the particle's position corresponding to the three previous scenarios .}
\end{figure}

Our experiment consists of a SiO$_2$ nanosphere with diameter of $\sim$100 nm levitated by a linearly polarized (the polarization orientation defines the $x$-direction) tightly focused Gaussian beam via an aspheric lens with a numerical aperture of 0.77. The focused beam forms an asymmetric trapping potential in the plane orthogonal to the optical axis and the trapping potential will be elongated in the direction parallel to the trapping beam's polarization. The focal asymmetry leads to different center-of-mass (COM) oscillation frequencies in the $x$ and $y$ directions, as illustrated in Fig. \ref{fig:one}(a). The position spectral density (PSD) of oscillations along the $x$ and $y$ directions at 6 mbar can be resolved in the measured photocurrent spectrum. We will refer to these oscillations as the $x$ and $y$ modes. The oscillation frequencies  are $\Omega_x = 2\pi \times 105$ kHz and $\Omega_y = 2\pi \times 120$ kHz. For small oscillation amplitudes, the two oscillation modes are independent and we can manipulate the energy in one mode by parametrically adjusting the trap stiffness by altering the trapping laser power at the nanoparticle's fundamental oscillation frequency or its first harmonic, while leaving the other mode unaffected. In this paper, the coupling between the two transverse oscillation modes is achieved by periodically rotating the asymmetric trapping potential in the transverse plane of the trapping beam\cite{frimmer2016cooling}. This is accomplished by rotating the linear polarization orientation of the trapping beam using an electro-optic modulator (EOM). The linear polarization orientation rotation, with respect to the $x$-direction, is described by $\phi(t) = \delta\cos{(\Omega t)}$, where $\delta$ is proportional to the voltage applied to the EOM. The modulation rotates the optical trapping potential as illustrated in Fig. \ref{fig:one}(b).

The dynamics of the motion of coupled mode levitated nano-particle can be expressed by following equations (Supplementary Section I):

\begin{align}
	\dot{\alpha}_{x}&=-\frac{{\rm i}}{2}\left(\left(\omega_{1}-\Omega\right)-{\rm i}\Gamma_{x}\right)\alpha_{x}+{\rm i}\beta_{x}\alpha_{y},\nonumber\\
	\dot{\alpha}_{y}&=\frac{{\rm i}}{2}\left(\left(\omega_{1}-\Omega\right)+{\rm i}\Gamma_{y}\right)\alpha_{y}+{\rm i}\beta_{y}\alpha_{x}
 \label{coupling equation}
\end{align}

\noindent where $\alpha_x$ and $\alpha_y$ is the particle oscillation amplitude in the $x$ and $y$ direction, $\omega_1 = \frac{\Omega_y^2-\Omega_x^2}{2\omega_0}$, and $\omega_0 = \frac{1}{2}\sqrt{\Omega_x^2+\Omega_y^2}$ is the carrier frequency. $\Gamma_x$ and $\Gamma_y$ are the damping or amplifying rate generated by the feedback intensity modulation or surrounding gas molecules and $\beta_{x}=\omega_{c}\sqrt{\Omega_{x}/\Omega_{y}}$, $\beta_{y}=\omega_{c}\sqrt{\Omega_{y}/\Omega_{x}}$, with $\omega_{c}=\left[\left(\Omega_{y}^{2}-\Omega_{x}^{2}\right)\delta\right]/(2\omega_{0})$ are coupling coefficients. According to the dynamics of $\alpha_{x}$ and $\alpha_{y}$, the mean phonon number of the two COM modes can be calculated: $\langle N_{i}\rangle=|\alpha_{i}|^{2}$,  with $i\in\{x,y\}$. The dynamics described by Eq. (\ref{coupling equation}) are analogous to the quantum treatment of a two-level atom driven by a classical light field \cite{frimmer2014classical} . The population of the two levels is denoted $\langle N_{x}\rangle$ and $\langle N_{x}\rangle$. The frequency splitting of the $x$ and $y$ mode is given by $\Omega_y-\Omega_x$. When the detuning factor $\Delta = \omega_1 - \Omega$ is sufficiently small,  the population of the two level system undergoes population oscillations at a generalized Rabi-frequency: $\Omega_R = \sqrt{\omega_c^2+\Delta^2}$. 

To demonstrate the coherent coupling of the two oscillation modes, see Fig. \ref{fig:one}(c), the COM oscillation in the $x$ direction is initially cooled to lower its energy via feedback before t = 0 ms. At t = 0 ms, feedback cooling on the $x$ mode is halted and the coupling between the $x$ and $y$ modes is activated, resulting in Rabi oscillations between the two oscillation modes.The phonon numbers of the two modes are proportional to the particle's oscillation amplitude in the two directions and determine the energy of each mode's COM motion. Figure \ref{fig:one}(d) illustrates the coherent control of the optical tweezer phonon laser and how the created coherence can be subsequently shared between each oscillation mode. The top row presents the two modes' oscillation amplitude and the bottom row is the position density map for the corresponding time traces.  The figure demonstrates the transition from a bimodal thermal state (0-100 ms) to single-mode thermal and single-mode coherent state (100-200 ms), and finally, following coherent mode coupling, the transition to two mode coherent state (200-300 ms).

\begin{figure}
\centering
\includegraphics[width=0.8\textwidth]{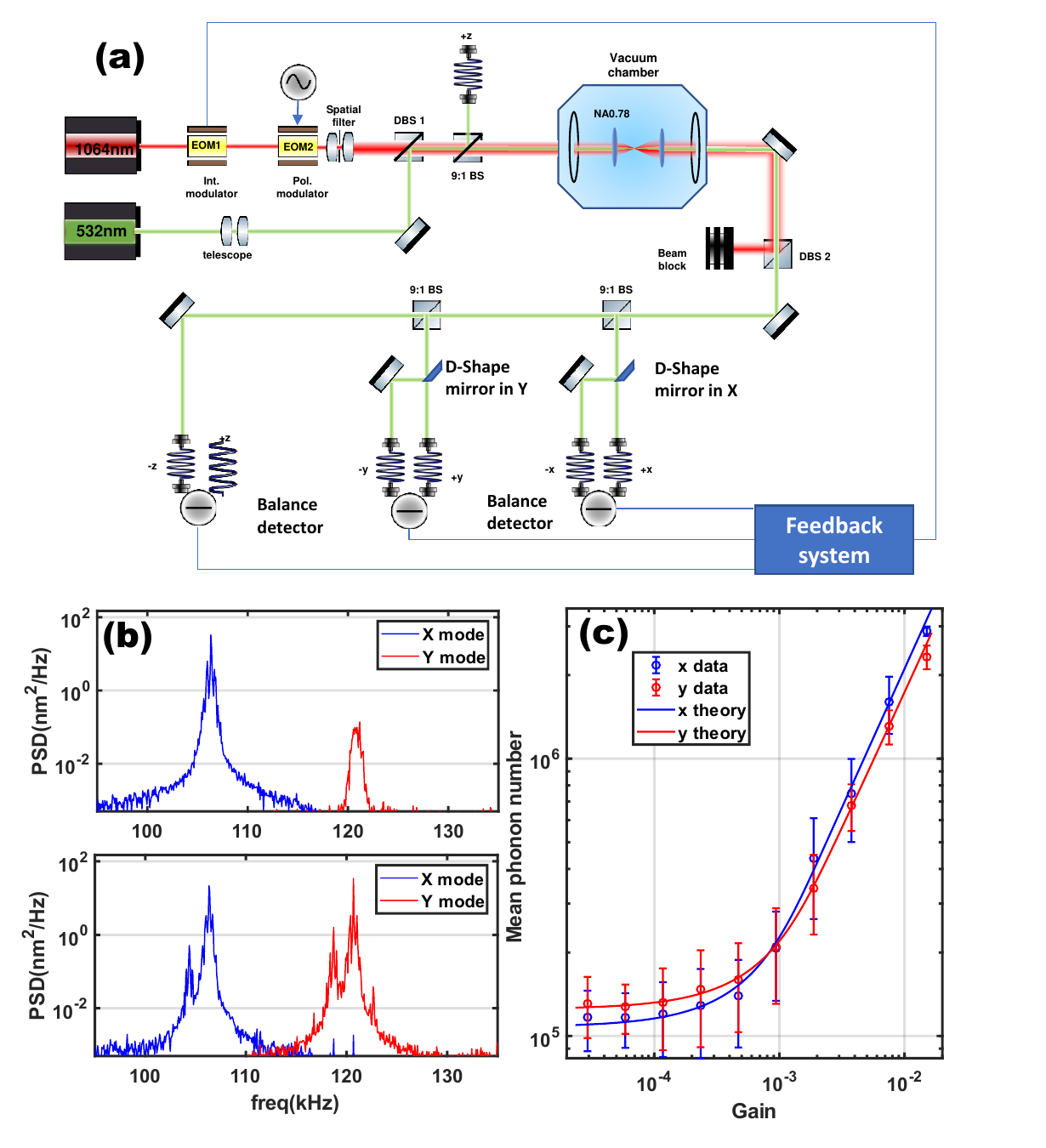}
\caption{\label{fig:two} (a) Experimental Setup. A 1064 nm trapping laser is used to trap the nanoparticle using a high NA (0.77) lens. Two electro-optic modulators (EOMs) are used to modulate the intensity and polarization of the trapping beam. A 532 nm probe beam is used to illuminate the particle and detect its motion. The motion in the $x$ and $y$ direction is detected by splitting the green beam into a left and right part ($x$ axis),a top and bottom part ($y$ axis) and then subtract the two using a balanced detector. The motion signal is sent to a feedback system and it generates a feedback signal to modulate the intensity of the trapping beam. (b) Position spectral density (PSD) of the $x$ and $y$ mode before (top panel) and after (bottom panel) coupling. (c) The mean phonon number in the two modes change with the gain in the $x$ mode when the two modes are initially coupled and feedback cooled.}
\end{figure}

Figure \ref{fig:two}(a) presents our experimental setup used to study the generation and sharing of coherence. The trapping beam is generated using a 1064 nm Nd:YAG laser, while a frequency doubled 532 nm laser serves as the probe beam. Both beams are focused by the same high NA aspheric lens. Subsequent to the vacuum chamber, the 1064 nm laser is blocked, and the 532 nm laser is collected to monitor the three-dimensional motion of the particle. For motion along the $x$ and $y$ axes, D-shaped half mirrors are introduced into the beams, positioned to separate the beams into right/left and top/bottom halves. The corresponding signal pairs are then directed towards a balanced detector, generating a difference current that solely retains the desired AC signal. Motion along the $z$ direction is monitored by subtracting the beam that traverses the chamber with a reference signal that is extracted from the probe beam before it enters the vacuum chamber. These three signals serve the dual purpose of recording motion characteristics of the particle and generating feedback signals to the electro-optic modulator 1 (EOM1) to modulate the intensity of the trapping beam for feedback heating or cooling. EOM2 is connected to an additional signal source to enable polarization modulation of the trapping beam.\newline

The motion signals are measured in units of voltage and our detectors are calibrated at a pressure of 6 mbar, where the COM temperature can be accurately approximated by room temperature (at atmospheric pressure the background gas is a room temperature bath). This enables the conversion of the voltage units into the displacement of the particle from the optical axis in nanometers. To characterize the properties of a phonon laser, we employ the mean phonon number to represent the energy of an oscillation mode. The mean phonon number is related to measurable particle displacement through the relationship: $\langle N\rangle = M\Omega_0\langle x^2\rangle/\hbar$, where $M$ is the mass of the particle, $\Omega_0$ is the oscillation frequency of the mode, $x$ is the displacement of the particle's center of mass with respect to the center of the trap, and $\hbar$ is Planck's constant.


\section{Results and Discussion}
Figure \ref{fig:two}(b) displays the steady state power spectral density (PSD) of the $x$ and $y$ modes before and after coupling, with the prerequisite that the $x$ mode is continuously lasing. Prior to the coupling, the $y$ mode exhibits a weak oscillation amplitude, while after coupling, the $y$ mode attains a comparable strength to that of the $x$ mode. Interestingly, both the $x$ and $y$ mode oscillation peaks exhibit a sidelobe to the left of the main lobe. The frequency difference between the main lobe and the sidelobe is identified as the Rabi-oscillation frequency, which results from the modulation of the motion of the two modes by the coherent polarization coupling. Figure \ref{fig:two}(c) depicts the variation of the mean phonon number for two modes with the gain modulation depth in the $x$ mode, while the coupling is consistently maintained. The gain modulation depth is defined as $M_a = \delta P_a/P_0$, where $P_0$ is the power of the trapping beam and $\delta P_a$ is the power modulation induced by the feedback amplification. The mean phonon number in the steady state can be found analytically (Supplementary Section I) and the solutions are plotted as solid curves in Fig. \ref{fig:two}(c).  It is observed that, beyond the threshold, the mean phonon number of both modes increases linearly with the gain applied to the $x$ mode and the threshold is almost the same for both the $x$ and $y$ mode when they are coupled. 

\begin{figure}
\centering
\includegraphics[width=0.8\textwidth]{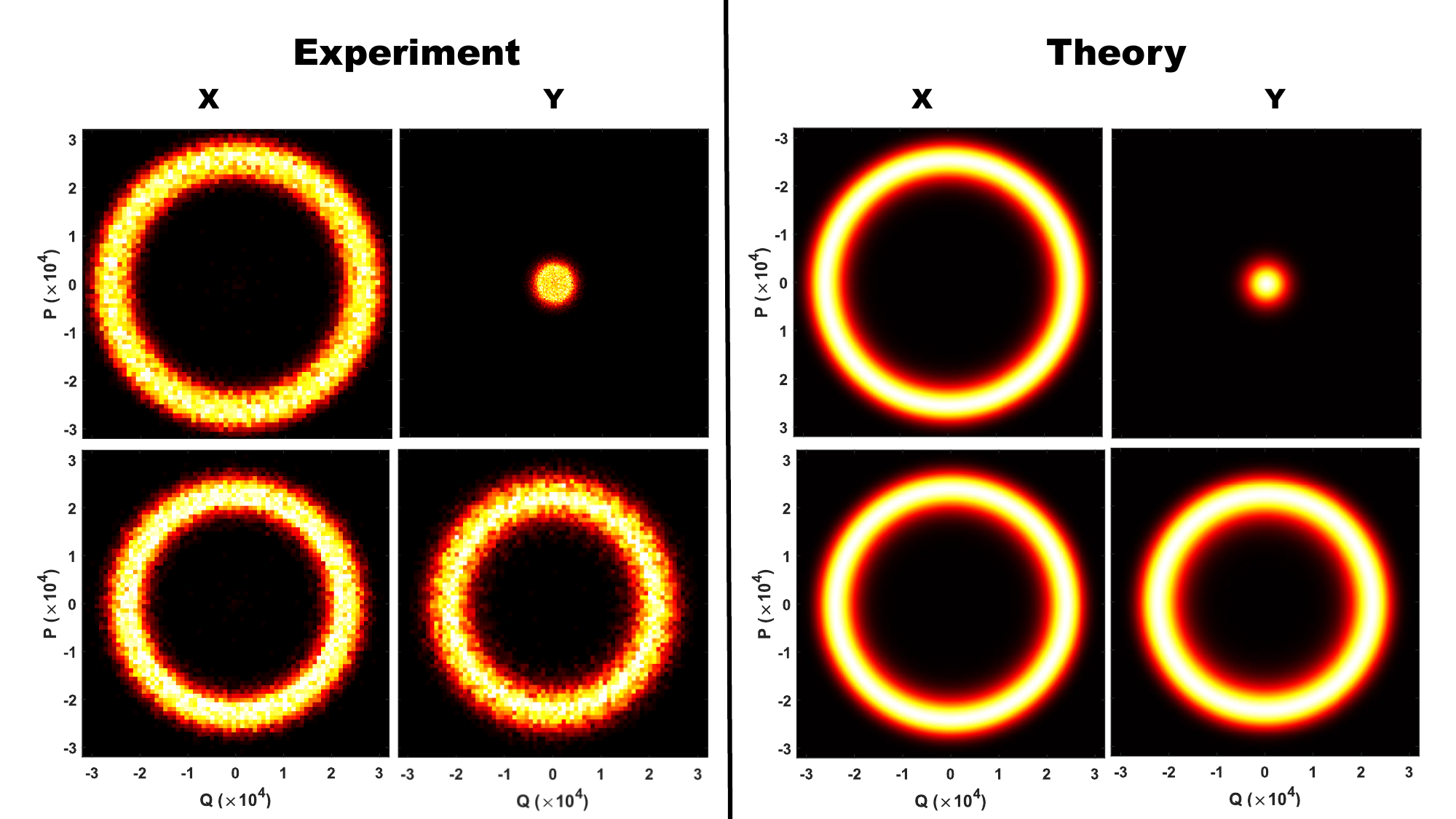}
\caption{\label{fig:three} Comparison of the measured quadratures of the oscillators' momentum-displacement distribution (left panel) to the theoretically expected phase space distribution (left panel) before (top row) and after (bottom row) coupling. A phase transition is observed in the $y$ mode after coupling.}
\end{figure}

For the $x$ and $y$ modes before and after the coupling, we investigate the steady state phase-space distribution of the oscillator by measuring the displacement and momentum of the particle and compare it with the theoretically expected oscillator's P-function representation (Fig. \ref{fig:three}). The left panel shows experimental data and the right panel shows the theoretical expectations. The axes are defined by the coordinates Q and P, where $Q = q/q_0$ and $P = p/p_0$ with $q_0 = \sqrt{\frac{\hbar}{2m\Omega_k}}$ and $p_0 = \sqrt{\frac{m\hbar\Omega_k}{2}}$ ($k = x,y$) are the in-phase and quadrature components of the motion scaled by the zero point position and momentum spread of the oscillator (Supplementary section I). While the $x$ mode is always in the coherent state, there is a clear transition from thermal state to coherent state in the $y$ mode before and after coupling.

\begin{figure}
\centering
\includegraphics[width=0.8\textwidth]{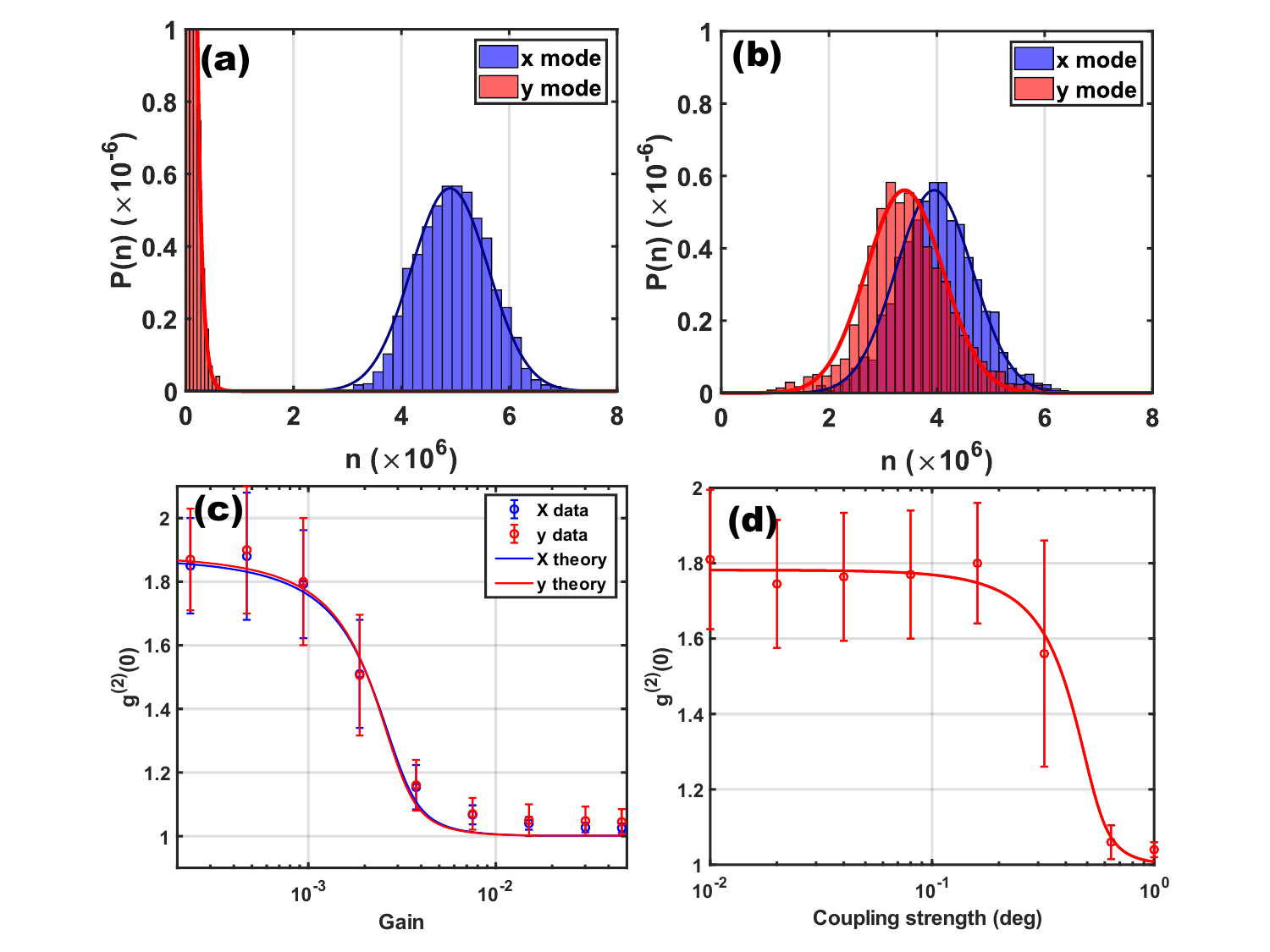}
\caption{\label{fig:four} The phonon number probability distribution before the coupling (a) and after the coupling (b). The phonon number histograms are well fitted by the Boltzmann distribution for a thermal state and the Gaussian distribution for coherent state. (c) The second order auto-correlation function $g^{(2)}_{(0)}$ of the phonon number in the $x$ and $y$ modes as a function of the gain in the $x$ mode when the two modes are continuously coupled and feedback cooled. Solid lines are the theoretical expectations. (d) $g^{(2)}_{(0)}$ of the phonon number in the $y$ mode changes with coupling strength when the $x$ mode is in a coherent state.}
\end{figure}

Figure \ref{fig:four} (a), (b) depict the steady state phonon number probability distribution in the $x$ and $y$ modes before (a) and after (b) coupling. Each histogram for the $x$ and $y$ modes is constructed from a 100 ms time section, with every phonon number measured in a 0.5 ms time window that is divided from the entire 100 ms data. Prior to the coupling, the $y$ mode displays a Boltzmann distribution with a low mean phonon number, indicating the $y$ mode is in a thermal state. In contrast, the particle's $x$ mode is prepared in the lasing (coherent) state and the phonon number probability distribution of the $x$ mode takes on a Gaussian distribution with relatively high mean phonon number. Following the coupling, both modes exhibit a Gaussian distribution and share a high value of the mean phonon number as compared to the thermal state, demonstrating that energy is transferred from the $x$ mode to the $y$ mode upon coupling. Like many optical lasers, we have a Gaussian limit of a Poissonian., which is predicted by our theoretical model (Supplementary Section II) and can be expressed by Eq. (\ref{marginal distribution r}):
\begin{align}
\phi(n_x)&=\mathcal{N}_{x}{\rm Exp}\left[-\frac{2}{A_{t}}3\gamma_{cx}n_{x}\left(n_{x}-\frac{1}{3\gamma_{cx}}\left(\frac{\Omega_{xy}^2}{4\gamma_{gy}}-\gamma_{gx}\right)\right)\right]\nonumber\\
\phi(n_y)&=\mathcal{N}_{y}{\rm Exp}\left[-\frac{2}{A_{t}}3\gamma_{cy}n_{y}\left(n_{y}-\frac{1}{3\gamma_{cy}}\left(\frac{\Omega_{xy}^2}{4\gamma_{gx}}-\gamma_{gy}\right)\right)\right],
	\label{marginal distribution r}
\end{align}
where $\mathcal{N}_{x(y)}$ is the normalization constant for the probability distribution of the $x$ ($y$) modes, $A_t$ is the photon scattering rate, $\gamma_{cx(y)}$ is the nonlinear feedback cooling rate in the $x$ ($y$) modes, $n_{x(y)}$ is the phonon number in the $x$ ($y$) mode, $\gamma_{gx(y)}$ is the gas damping rate in the $x$ ($y$) mode.  $\Omega_{xy}=\frac{\left(\Omega_{y}^2-\Omega_{x}^2\right)}{2\sqrt{\Omega_{x}\Omega_{y}}}\delta$ is the coupling between the $x$ and $y$ modes under the rotating angle modulation.

In Fig. \ref{fig:four}(c) and (d) we examine the steady state statistics of the system in more detail and show the second-order phonon auto-correlation function at zero time delay, $g^{(2)}(0)=(\langle N^2\rangle-\langle N\rangle)/\langle N\rangle^2$, where $\langle N^2\rangle$ is the second moment of the distribution. Note that $1<g^{(2)}(0)<2$ for a thermal state and $g^{(2)}(0)=1$ for a coherent state \cite{gerry2005introductory}. Experimental phonon number distributions were measured by recording oscillator dynamics in a time window of 100 ms for each point in the plots. The length of the error bar represents $\pm1$ standard deviation of 100 such measurements. Figure \ref{fig:four}(c) illustrates a gradual change of $g^{(2)}(0)$ for the $x$ and $y$ modes from a thermal state to a coherent state with respect to the $x$ mode gain when both modes are initially coupled and feedback cooled. The $g^{(2)}(0)$ function evolution for the coupled $x$ and $y$ modes are almost the same. In Fig. \ref{fig:four}(d), while the $x$ mode is lasing, we plot the $g^{(2)}(0)$ of  the $y$ mode as a function of coupling strength $\delta$.  The data and theory evidence a threshold for coupling strength that needs to be exceeded to ensure the two modes couple coherently.

\begin{figure}
\centering
\includegraphics[width=0.8\textwidth]{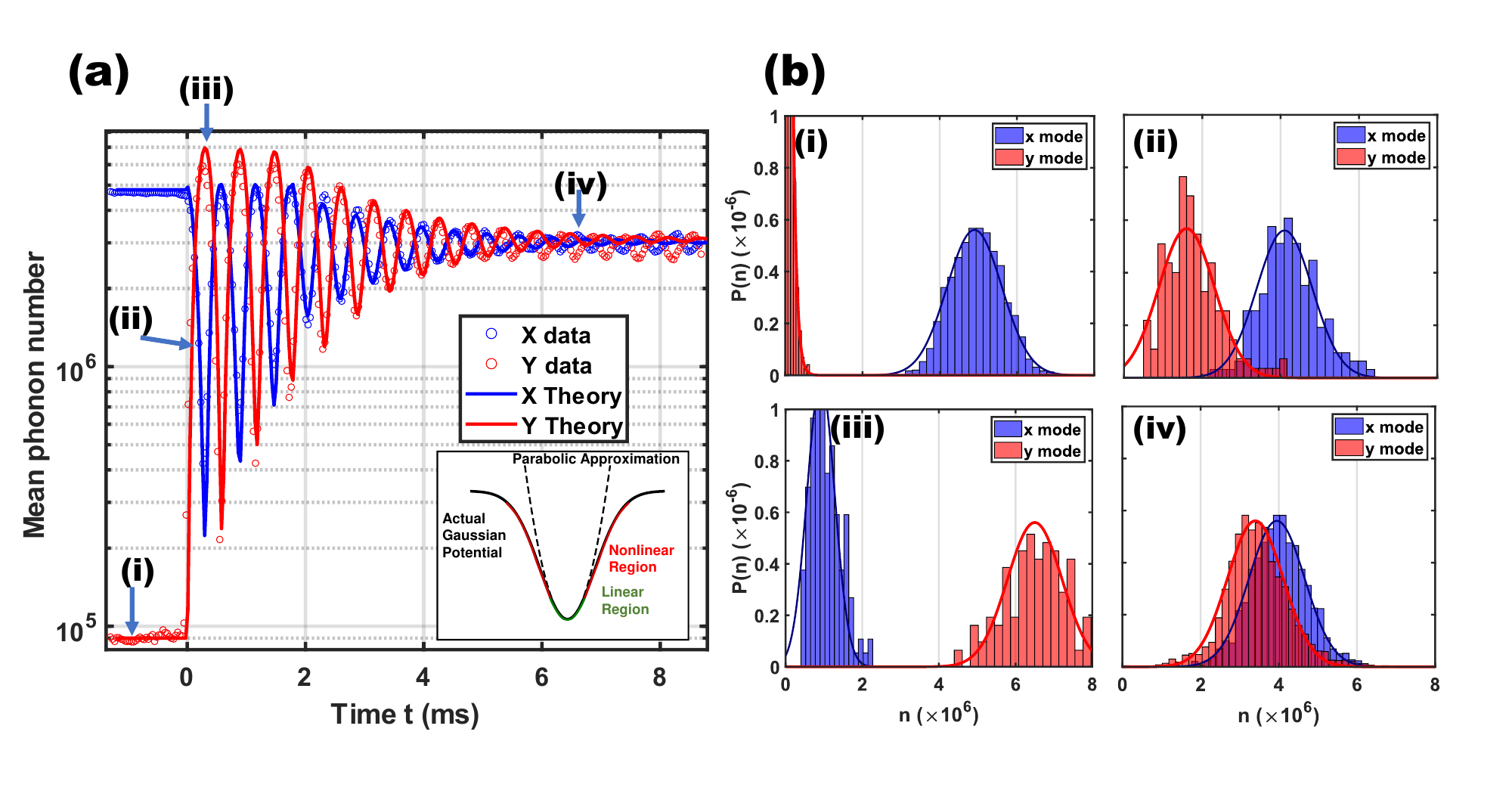}
\caption{\label{fig:five} (a) The transient phonon number exchange in a coupled mode optical tweezer phonon laser after the coupling is switched on at 0 ms. Small circles are experimental data, solid lines are the numerical solution of the coupled mode equations; Eq. (\ref{coupling equation}). The inset illustrates the linear region and nonlinear region of the trapping potential (b) The four figures on the right show the phonon number probability distribution at four time points (i), (ii), (iii), and (iv) labeled on the left curve.}
\end{figure}

The transient dynamics of the coupled mode phonon laser system, as the coupling is turned on at t = 0 ms, are explored in Fig. \ref{fig:five} (a). Each measurement spans 10 ms and between measurements we terminate the coupling for 100 ms letting the $y$ mode cool down to the lower energy level by feedback cooling which reinitializes the particle's state. By repeating this process 100 times, statistical distributions of the phonon population of the oscillator can be constructed as the system evolves and the mean value can be computed from these distributions.  The experimental data are plotted using small circles, and the solid lines are acquired by numerically solving the coupled mode equation Eq. (\ref{coupling equation}) under the condition of setting  $\Gamma_{i}=2(\gamma_{g}-\gamma_{a}\delta_{i,x}+6\gamma_{ci}|\alpha_{i}|^{2})$ with $i\in\{x,y\}$.
Upon activation of the coupling, the two modes initially exhibit Rabi-oscillation, a hallmark of their coherent coupling. Subsequently, the mean phonon number of the $y$ mode increases and remains constant, matching the level of the $x$ mode. At the outset of the coupling, the data conforms well to the theoretical curve, however, once both modes attain a steady state, the theory predicts constant energy levels for both modes while the experimental results show a minor energy transfer between them. This deviation might arise from disturbances in the vacuum chamber which could potentially impact the energy difference between the two modes, leading to continuous energy transfer even at steady state. 
Furthermore, towards the end of the oscillation process, a slight frequency mismatch between the experimental and theoretical results is observed. This discrepancy may arise from the fact that when the $y$ mode is driven into the lasing state, its oscillation amplitude reaches the nonlinear region of the trapping potential as illustrated in the inset of Fig.\ref{fig:five}(a). Specifically, the area within $\frac{1}{10}$ of the potential width, approximately 60 nm in our case, can be regarded as the linear region, while the region beyond exhibits nonlinear effect where the trapping potential well deviates from the parabolic approximation. Consequently, the oscillation frequency of the $y$ mode shifts to lower frequency. As the modulation frequency remains fixed, this change in detuning factor leads to variations in the Rabi-oscillation frequency. The four panels in Fig. \ref{fig:five}(b) illustrate the probability distribution of the phonon numbers at four different time points (i), (ii), (iii), and (iv), which are labeled in Fig. \ref{fig:five}(a). At points (ii) and (iii), there are not enough data points to calculate the probability distribution for such a small time window. Therefore, at these two points, 100 measurements are considered simultaneously, and the phonon probability density is plotted through all 100 data sets. The $y$ mode evolves from a thermal Boltzmann distribution in case (i) to a Gaussian distribution in case (iv). Cases (ii) and (iii) represent intermediate states when the two modes are continuously exchanging energy. In case (iii), the mean phonon number in the $y$ mode is greater than that in the $x$ mode, even thought there is no heating applied to the $y$ mode which is consistent with the oscillation behavior shown in Fig. \ref{fig:five} (a).

\begin{figure}
\centering
\includegraphics[width=0.8\textwidth]{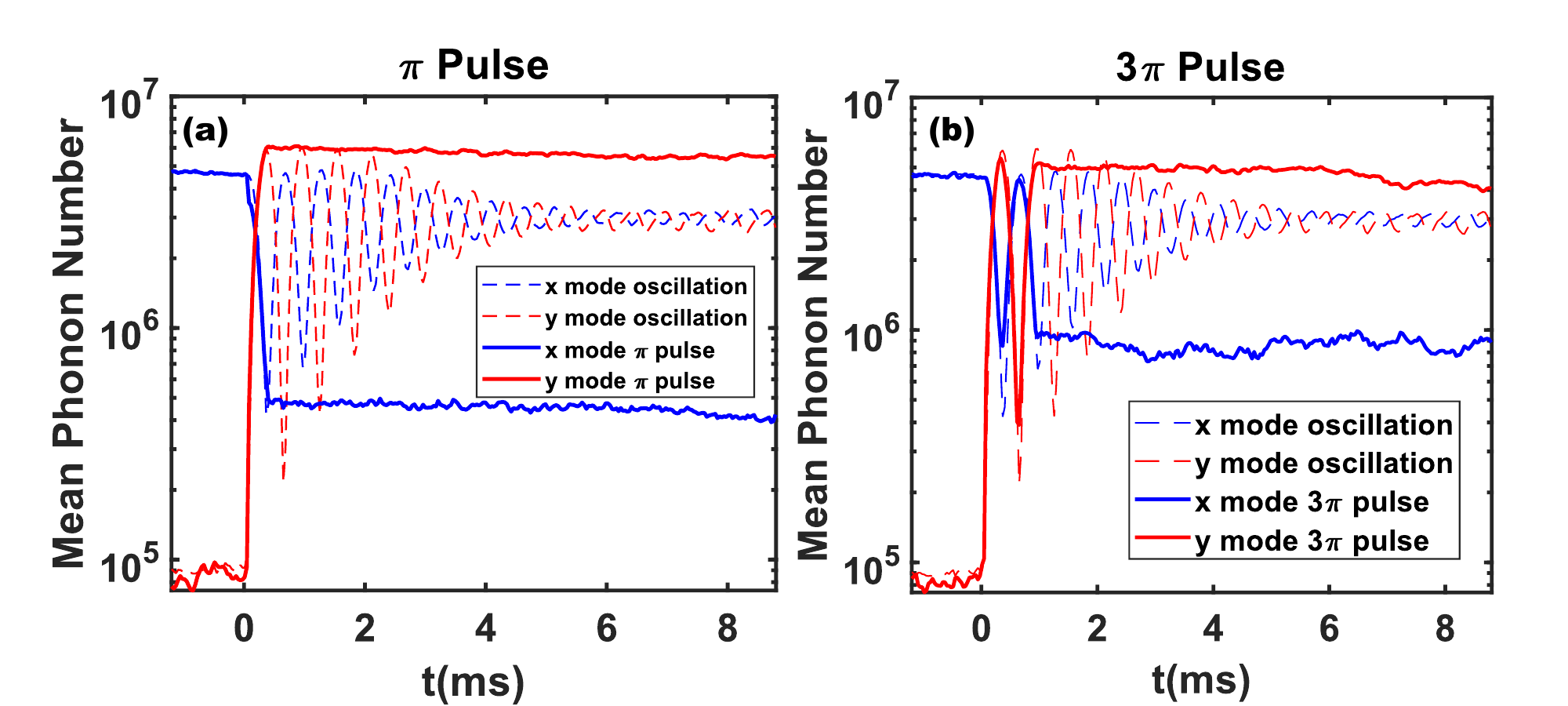}
\caption{\label{fig:six} (a): The $x$ and $y$ mode dynamics upon application of a $\pi$ coupling pulse at t = 0 ms. (b): The $x$ and $y$ mode dynamics upon application of a 3$\pi$ coupling pulse at t = 0 ms.  Dashed curves in the background are continuous coupling shown in Fig. \ref{fig:four}.}
\end{figure}

The coherent coupling also makes possible coherent control of the phonon laser dynamics. In the previous experiments, the coupling is turned on and kept on after t = 0 ms. After half a Rabi-oscillation period, the $y$ mode has a much larger mean phonon number than the $x$ mode (point (iii) in Fig. \ref{fig:five}). It is possible to stop the coupling at this time point and let the system stay at its new energy. This is achieved by using pulsed coupling. Two coupling pulses, $\pi$ and $3\pi$ respectivelty, are presented in Fig. \ref{fig:six} as solid lines, which correspond to half and one and a half Rabi-oscillation periods. The dashed line represents continuous coupling for comparison. It should be noted that these results are still averages over 100 measurements, and different particles show similar results for coupling pulse experiments. The present demonstrations confirm the possibility of coherently controlling the optical tweezer phonon laser dynamics.

\section{Conclusion}
In this paper, we have introduced a coupled mode optical tweezer phonon laser. By examining the steady state mean phonon number, the phase space distribution, the second order auto-correlation function and transient dynamics, we theoretically and experimentally confirmed the transition from a thermal state to a coherent state for the $y$ mode oscillation even without feedback amplification applied to that mode. The results of our study demonstrate that coherence creation isn't exclusively dependent on feedback cooling and amplification, it can also be created in a thermal mode by coherently coupling that mode to one that is already in a coherent state. Outside of the ability to control the statistical properties of the optical tweezer phonon laser, by coupling multiple modes it should be possible to create two-mode squeezed states \cite{pontin2016dynamical,li2015generation,woolley2014two,han2020generation}, evade back action in measurement \cite{hertzberg2010back,ockeloen2016quantum} and, in the quantum regime \cite{delic2020cooling,tebbenjohanns2021quantum} explore entanglement  \cite{li2015generation}. Further, by carefully tuning the gain, loss and coupling of two modes, one can drive the system into the PT-symmetry regime \cite{sharma2022pt}. Finally, these experiments are the first steps toward multimode and multiparticle transfer of coherent phonons.

\begin{backmatter}

\bmsection{Acknowledgments}
M. B. and A. N. V. acknowledge support from ONR Grant N00014-18-1-2370 . A. N. V. acknowledges support from a University of Rochester Research Award. S. S. acknowledges funding from National Research Foundation of Korea (NRF) Grant No. NRF-2022R1I1A1A01053604 and BK21 postdoctoral fellowship from KAIST

\bmsection{Disclosures}
The authors declare no conflicts of interest.

\bmsection{Data Availability Statement}
Data underlying the results presented in this paper are not publicly available at this time but may be obtained from the authors upon reasonable request.

\bmsection{Supplemental document}
See the supplement for supporting content. 

\end{backmatter}

Prism interface.




\end{document}